\documentclass[sigconf]{acmart}

\usepackage{url}
\usepackage{hyperref}
\usepackage{svg}
\usepackage{tikz}
\usepackage{verbatim}
\usepackage{caption}
\usepackage{breqn}
\usepackage{footnote}
\usepackage{subcaption}
\usepackage{tablefootnote}
\usepackage[font={small}, belowskip=-5pt,aboveskip=5pt]{caption}
\usepackage{cleveref}
\usepackage{tabularx}
\usepackage{multirow}
\usepackage{array}
\usepackage{enumitem}
\usepackage{algorithm}
\usepackage{algorithmic}

\usepackage{fancyvrb}
\usepackage{fvextra}

\crefname{section}{§}{§§}
\Crefname{section}{§}{§§}

\AtBeginDocument{%
  \providecommand\BibTeX{{%
    \normalfont B\kern-0.5em{\scshape i\kern-0.25em b}\kern-0.8em\TeX}}}

\copyrightyear{2022}
\acmYear{2022}
\setcopyright{licensedcagov}\acmConference[SAC '22]{The 37th ACM/SIGAPP Symposium on Applied Computing}{April 25–29, 2022}{Virtual Event, }
\acmBooktitle{The 37th ACM/SIGAPP Symposium on Applied Computing (SAC '22), April 25–29, 2022, Virtual Event, }
\acmPrice{15.00}
\acmDOI{10.1145/3477314.3507097}
\acmISBN{978-1-4503-8713-2/22/04}

\begin{document}


\title{ANUBIS: A Provenance Graph-Based Framework for Advanced Persistent Threat Detection }

\author{Md. Monowar Anjum, Shahrear Iqbal}
\authornote{Corresponding Author}
\affiliation{%
  \institution{National Research Council}
   \city{Fredericton}
   \state{NB}
   \country{Canada}
}
\email{{mdmonowar.anjum, shahrear.iqbal}@nrc-cnrc.gc.ca}

\author{Benoit Hamelin}
\affiliation{%
 \institution{Tutte Institute for Mathematics and Computing}
  \city{Ottawa}
  \state{ON}
  \country{Canada}
}
\email{benoit.hamelin@cyber.gc.ca}


\begin{abstract}
 We present ANUBIS, a highly effective machine learning-based APT detection system. Our design philosophy for ANUBIS involves two principal components. Firstly, we intend ANUBIS to be effectively utilized by cyber-response teams. Therefore, prediction explainability is one of the main focuses of ANUBIS design. Secondly, ANUBIS uses system provenance graphs to capture causality and thereby achieve high detection performance. At the core of the predictive capability of ANUBIS, there is a Bayesian Neural Network that can tell how confident it is in its predictions. We evaluate ANUBIS against a recent APT dataset (DARPA OpTC) and show that ANUBIS can detect malicious activity akin to APT campaigns with high accuracy. Moreover, ANUBIS learns about high-level patterns that allow it to explain its predictions to threat analysts. The high predictive performance with explainable attack story reconstruction makes ANUBIS an effective tool to use for enterprise cyber defense.
\end{abstract}

\ccsdesc[500]{Security and privacy~Intrusion detection systems}

\keywords{Advanced Persistent Threat, Provenance Graph Analysis}


\maketitle

\section{Introduction}

The term \textbf{Advanced Persistent Threat} (APT) was first introduced in reference to intrusion into secure systems by US Air Force in 2006~\cite{Hassannataj_20}. Today, the term APT is used for a wide range of cyber attacks executed over a long period of time to gain access to highly confidential information or to compromise high-integrity resources while remaining undetected. According to the FireEye-Mandiant M-Trends Threat Intelligence Report 2021, there are currently 41 active APT groups in activity today~\cite{fireeye2020report}. These groups are responsible for targeted campaigns such as APT29's attempt to steal COVID-19 vaccine-related information \cite{apt29covid}. Another prominent example is 2020's United States federal government data breach via the supply chain exploitation of the SolarWinds enterprise software vendor, which was orchestrated by unknown state-sponsored actors \cite{solarwinds2020}. Cyber security trend reports suggest that APT actors are taking advantage of newly discovered zero-day vulnerabilities in popular software to gain access into safety-critical systems~\cite{infosys2021report, symantec2011advanced}. 

Traditional threat detection systems are not suitable for detecting long running APT campaigns. Signature-based detection systems perform poorly in the face of zero-day exploits, polymorphic malware and living-off-the-land tactics. On the other hand, systems based on anomaly detection fail to model long running network behavior due to resource constraints of storing and processing high-density telemetry and log events for a long period of time~\cite{feng2003anomaly,maggi2008detecting, sekar2000fast, somayaji2000automated, mutz2007exploiting}. Moreover, systems that take long term behaviour into account limit their analysis to paired call stack event occurrences~\cite{long_modeling_system_2017}.

Recent works ~\cite{Milajerdi_2019, han2020unicorn, milajerdi2018propatrol, barre2019mining} suggest that system provenance graphs are a more effective data source for APT detection. A system provenance graph is a directed acyclic graph (DAG) that represents causal relationships between running processes and objects (e.g., files, network flow, threads) in a system. It can connect events that are temporally distant but causally related~\cite{pasquier2017practical}. This property of system provenance graphs provides rich contextual information regarding an event's neighborhood and it's parent event. Leveraging this property leads to robust separation of benign and malicious events.\\

\noindent \textbf{Motivation.} Prior research works that used provenance graphs for APT detection suffer from two major limitations. First, provenance graph-based systems that use graph edge matching rules are very sensitive~\cite{Milajerdi_2019, milajerdi2019poirot}. APTs can evade these systems by exploiting new zero-day vulnerabilities. Second, in memory analysis of system provenance graph is prohibitively expensive~\cite{manzoor2016fast, Milajerdi_2019} as provenance graph grows exponentially with time. In our work, we mitigate the first limitation by learning APT attack graph patterns from provenance data. This allows our system to learn high level structures of APT attacks. We believe that these structures are more reliable indicators of attacks than low level handcrafted edge matching rules. We circumvent the second limitation by proposing a novel graph neighborhood encoding method. Our proposed method uses Poisson distribution to encode the neighboring events in a fixed length vector. This allows our system to efficiently perform in-memory provenance graph analysis during runtime.

Prior research works that used provenance graph for APT attack reconstruction focus on storytelling over explainability~\cite{alsaheel2021atlas, hossain2017sleuth}. Let us illustrate this by an example. Suppose, a \verb|cmd.exe| process opens a word document by triggering \verb|winword.exe|. This is a benign action. The opposite sequence where \verb|winword.exe| is triggering a \verb|cmd.exe| is considered malicious action. Current research works focus on telling this story where they point out to the cyber analyst that a suspicious activity took place and the story is: \verb|winword.exe| triggered \verb|cmd.exe|. However, current works can not answer if the cyber analyst asks ``\textbf{why} this activity is suspicious?''. The most conclusive way for an APT detection system to answer that question is to map any detected suspicious activity to the most similar activity in it's training dataset. In the above example, the answer given by the APT detection system should be: ``This activity represents maximum similarity with an macroless shellcode injection example in training set.'' This approach of attack story reconstruction provides explainability during the attack investigation. Moreover, it also provides an insight into the detection system's health (i.e., if the system is making majority false positive predictions for a specific type of events). \textit{In this work, we explore the explainable approach for APT attack reconstruction. To the best of our knowledge, this is the first work that takes explainability into account for attack story reconstruction.}  

\noindent \textbf{Contributions.} We propose ANUBIS, a supervised machine learning approach for detecting APTs from provenance graph data and explaining the predictions to cyber-threat responders. ANUBIS is trained on event traces (a sequence of events related by parent-child relationship) which are generated by walks on the provenance graph. The model used for training ANUBIS is a Bayesian Neural Network (BNN). During the operation phase, ANUBIS predicts the class of the event traces that are generated from the streaming provenance graph. These predictions are also associated with an uncertainty score provided by our BNN model. If the prediction has low uncertainty score, ANUBIS provides explanation of the prediction by matching with the most similar example in the training set. 
In summary, we make the following contributions:

\begin{itemize}
    \item We propose a provenance graph-based supervised APT detection framework.
    \item We develop a novel graph-neighborhood encoding method using Poisson distribution that greatly reduces the memory footprint of the graph-based detection system.
    \item We provide explanation of the predictions according to a strategy that depends on the uncertainty of the prediction given by the BNN model.
    \item We evaluate ANUBIS on a recent APT dataset called Operationally Transparent Cyber (OpTC) released by DARPA. ANUBIS achieves an accuracy of 99\% with precision over 98\% and false positive rate of less than 2\% which show the efficacy of our approach.

\end{itemize}

\section{Background}
\subsection{APT Life Cycle}

According to the APT life cycle model~\cite{Milajerdi_2019}, a typical APT attack consists of the following steps: (1) Initial Compromise; (2) Foothold Establishment; (3) Privilege Escalation; (4) Internal Reconnaissance; (5) Lateral Movement; (6) Achieve Persistence; and (7) Complete mission. Let us consider the 2020 US federal government data breach as an example. The adversary initially compromised the system by implanting a malicious dll (SUNBURST) in the Solarwinds Orion product. Then the adversary communicated with a number of malicious domains via DNS queries (CNAME records) and established foothold in the compromised system. Afterwards, the adversary used Mimikatz to harvest credentials in order to escalate their privilege. The internal reconnaissance was performed by periodic scanning of shared file system. The adversary then moved laterally to spread into other hosts. In order to achieve persistence, the adversary moved away from the original source of compromise (malicious dll) and leveraged VPN vulnerabilities to gain continuous access to the system. Once the adversary detached itself from the source of initial compromise, it became very difficult for cyber-threat responders to detect and track the adversary activities. Lastly, the adversary exfiltrated confidential data by common protocols such as HTTPS and SMB from the hosts where the adversary successfully achieved persistence~\cite{fireeye2020report}. 
\subsection{Challenges of APT Detection}
Any provenance graph-based APT detection system will have to overcome three main challenges that can influence the philosophy of the system design. One of them is related to the nature of the APT attack, another is related to the log data which is used by the system to construct the provenance graph and the final one is related to the time sensitive nature of the task of APT detection. 

\textbf{First}, APTs maintain a very low profile in the compromised system. Their activities blend with normal system execution trace. These long running, stealthy, and persistence attack techniques make the task of detecting APT very difficult~\cite{alshamrani2019survey}. 

\textbf{Second}, Enterprise system hosts can produce terabytes of event log data on a daily basis which are used to construct the provenance graph. APT activities are usually an extremely small percentage of this data (typically less than 0.001\%). Therefore, designing a detection system for APT which has low false positive rate is difficult~\cite{manzoor2016fast}. High volume of false positive alarm can cause ``Threat Fatigue" among system administrators which can lead to dismissal of threat alerts~\cite{han2020unicorn}. 

\textbf{Third}, detection of long-running APT campaign is a time sensitive task. The longer an APT is within a system, the more potential damage it can inflict. Therefore, it is imperative that APT detection systems should isolate and summarize the suspicious events quickly and present them to the cyber-response team for triage.

\subsection{Bayesian Neural Network} Bayesian Neural Network (BNN) is a special class of neural network proposed by Blundell \textit{et al.}~\cite{blundell2015weight}. Ordinary Neural Networks (NN) learn fixed weights and biases from the training data to approximate the function for the given task. However, a BNN learns a \textbf{distribution} of weights and biases to approximate the function for a given task. This property allows BNNs to determine uncertainty in its prediction. Let us illustrate this with an example. Consider a NN which is trained on the MNIST dataset~\cite{lecun2010mnist} to classify grayscale images of integer numbers between 0 to 9. After training, the NN will achieve good accuracy on a test set of grayscale images containing digits only. If we ask the same NN to predict the class of a grayscale image that contains a car, the network will still predict the class of a digit. This is because the NN is trained to approximate a function that maps input images to classes of digits. If the input image is not a digit, the NN does not know how to handle that.

 Let us consider a BNN in a similar scenario. After being trained on the MNIST dataset, the BNN will predict the class of grayscale images containing digits with very low uncertainty score. In other words, the BNN is \textbf{confident} that it is predicting the right class for that image. Now, if the BNN is asked to predict a grayscale image containing a car, it will classify the car as a digit. However, it will attach a very high uncertainty score with this prediction output. This means the BNN is \textbf{not confident} in its prediction. This particular capability to attach an uncertainty score is extensively leveraged in ANUBIS design (Section \ref{sec:anubis_design}).
 

\subsection{Problem Statement}
The problem that is addressed in this work is the detection and explanation (regarding the detection) of APT attacks from the system provenance graph. An ideal detection system that can tackle the problem should have the following properties:
\begin{itemize}
    \item Ability to continuously monitor the system events and construct the provenance graph of the system.
    \item Capacity to detect events that does not conform to the known benign event patterns and provide a detailed explanation to the cyber-response team for further actions. Explanations can include matching the event with a known malicious event pattern, attack story reconstruction, etc.
\end{itemize}

We also make the following assumptions: 
\begin{itemize}
    \item The adversary can not compromise the provenance graph used to generated event traces for training ANUBIS.
    \item The adversary can not compromise the event logging mechanism of the host system during operational phase. This assumption is central to the ANUBIS design. 
    \item Malicious event traces show sufficiently different behavior from benign event traces. 
\end{itemize}
\begin{figure}[t]
    \centering
    \includegraphics[width=0.45\textwidth]{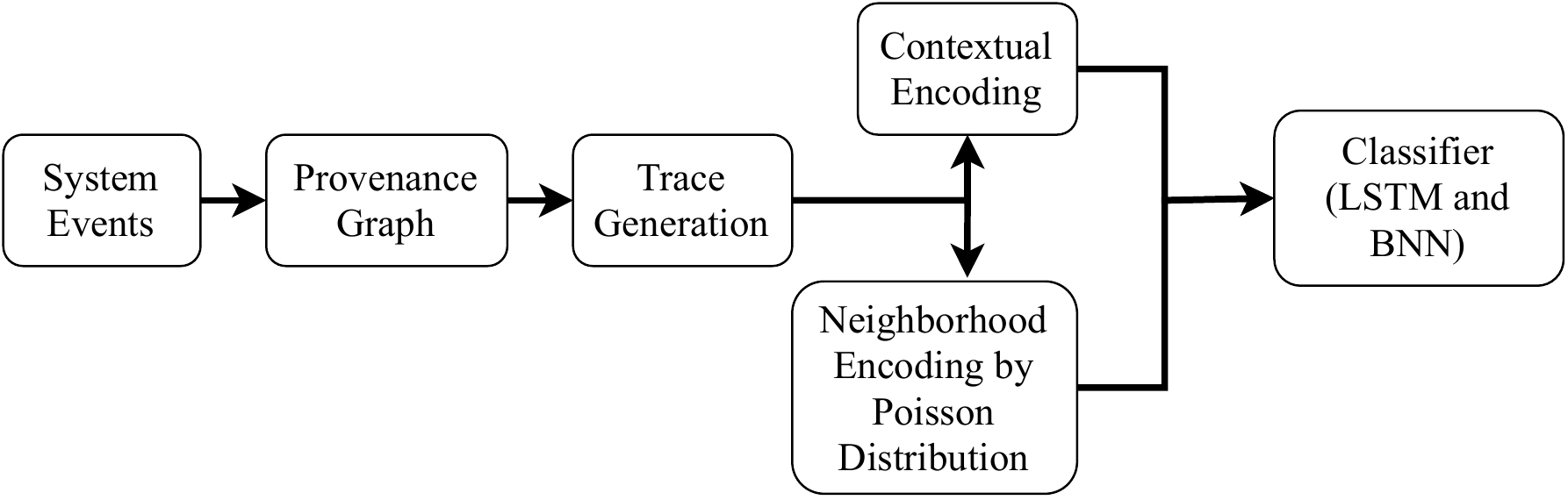}
    \caption{Block diagram showing the steps of ANUBIS training.}
    \label{fig:anubis_training}
\end{figure}
\section{ANUBIS Design}
\label{sec:anubis_design}
ANUBIS is a host-based system which is trained on event-traces generated from provenance graphs from APT infected systems. It has a two-phase design: Training and Operation. Figure~\ref{fig:anubis_training} shows ANUBIS training phase design.

\subsection{Training Phase Design}
\label{subsec:training}
\subsubsection{Provenance Graph Construction.} Provenance graph is a directed acyclic graph that can causally connect system events even when they are temporally distant. During the training phase, ANUBIS takes the system events generated by the logging mechanism and constructs a provenance graph. Each internal node of the provenance graph represents a process creation event. The children of the internal nodes represent the events where the process interacted with a system resource, i.e., open/modify/delete a file, send a packet over the network, launch or communicate with another process, etc. This structure of provenance graph provides valuable information regarding the context, causality and neighborhood of a particular system event.

\begin{figure}[t]
    \centering
    \includegraphics[width=0.45\textwidth]{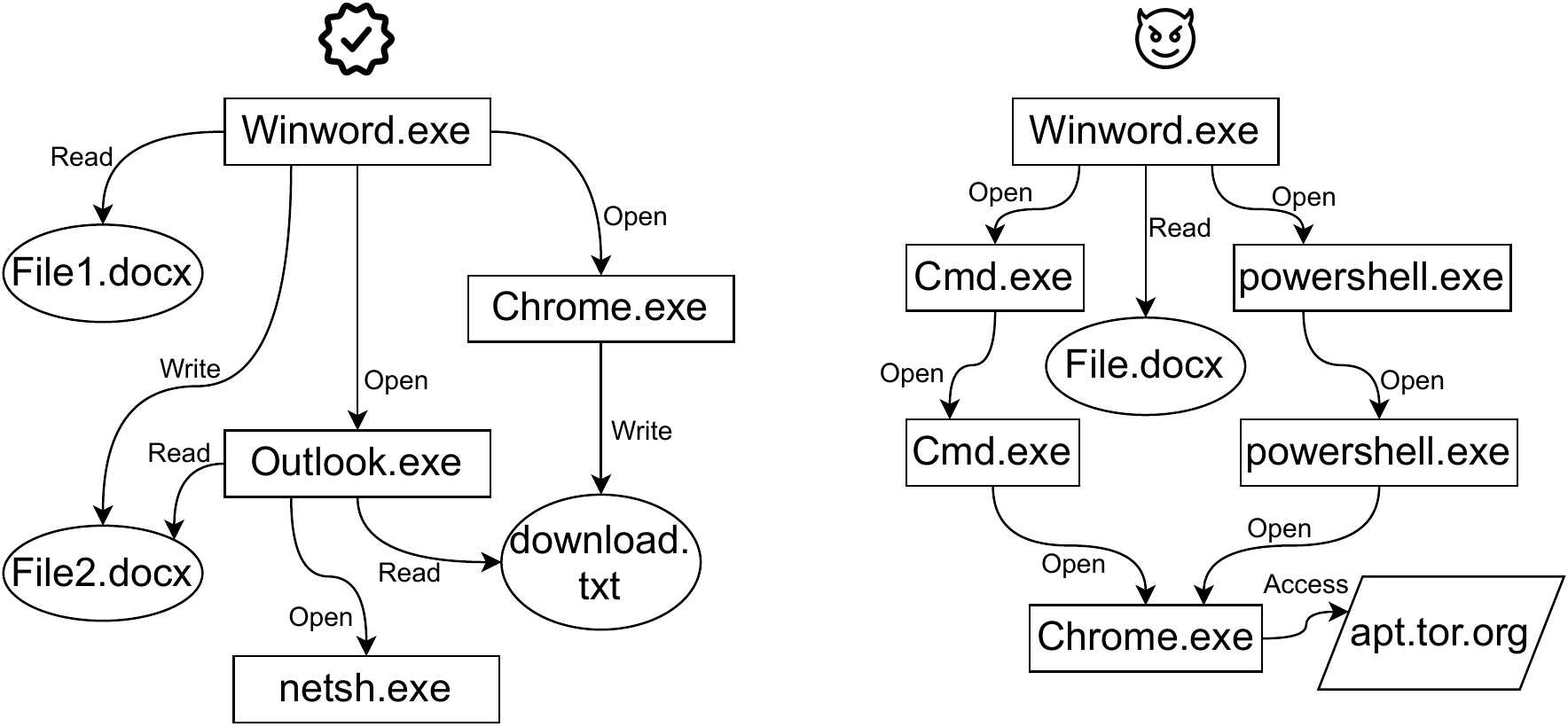}
    \caption{Two subgraphs of the system provenance graph. The left one shows a benign subgraph and the right one shows the malicious subgraph of the same process.}
    \label{fig:benign_and_malicious graph}
\end{figure}

Let's illustrate this with an example. Figure \ref{fig:benign_and_malicious graph} shows the subgraph of the system provenance graph for a benign instance and a malicious instance of the MS Word process. The context and causal relationships for events are different between benign and malicious subgraphs. Additionally, neighborhood of events in the subgraphs is different from each other. These properties help ANUBIS to categorize benign and malicious event traces accurately. 

\subsubsection{Event Trace Generation.} ANUBIS generates event traces from the system provenance graph for training the BNN. An event trace is defined as a sequence of events who are related by parent-child relationships. Formally speaking, let's select a random internal node on the provenance graph and consider a walk of length $l$ which starts on that node. This walk will generate an event sequence $[e_1, e_2, e_3,\cdots,e_l]$. Here $e_1$ is the parent of $e_2$, $e_2$ is the parent of $e_3$ and so on. This sequence of events is called an event trace. For example, \textit{winword.exe $\rightarrow$ chrome.exe $\rightarrow$ download.txt} is an event trace from the benign subgraph of Figure \ref{fig:benign_and_malicious graph}.


\subsubsection{Encoding Event Traces into Floating Point Vectors.} A neural network or any machine learning algorithm for that matter, takes floating point vectors as input. Therefore, ANUBIS converts the generated event traces into floating point vectors that can be fed into neural network layers. ANUBIS encodes two types of information from the provenance graph. The first one is causal and contextual information and the second one is neighborhood information.


\paragraph{Encoding Causal and Contextual Information}
\label{encoding1}
Prior works~\cite{han2020unicorn,wang2020you} have shown that one of the key inputs for any APT detection model is the causality and contextual information. For an event trace of length $l$, we encode the causality and contextual information of that trace in a $l\times d_1$ matrix. Each event in the trace corresponds to a $d_1$-dimensional row vector in the matrix. The row vector contains encoded information about the event type, time difference from the parent event, the name and location of the parent process, the type of the process responsible for triggering the process (system or user), etc. We do not encode ephemeral properties of events in the vector. Ephemeral properties include properties that change across different instances of the same process (e.g., \verb!pid! , \verb!ppid!, and \verb!sid!). We exclude these properties to avoid generating noisy input for ANUBIS training.

\paragraph{Encoding Neighborhood Information}
\label{encoding2}
Another key input for any APT detection model is the neighborhood information of an event. For example, Figure \ref{fig:benign_and_malicious graph} shows that the benign provenance subgraph has different neighboring event distribution compared to the apt-infected provenance subgraph. We encode the neighborhood event distribution in a floating point vector. 


Now we define neighborhood of an event and describe the specific distribution that is used in this work. The neighborhood of an event consists of the siblings that occurred prior to the event. In order to model the neighborhood of an event, ANUBIS considers every new event generation as a binary random variable $X\in\{0,1\}$. Empirical observation suggests that $X$ is rarely 1\footnote{New event generation in the neighborhood of a provenance graph node is a rare occurrence. Thousands of events are generated every second in the system. However, not all of them are generated by the same process. Therefore, they end up in different neighborhoods of the graph}. Consequently, we need to choose a distribution that can model rare event occurrences. One such distribution is Poisson distribution which is widely used in the literature to model rare events (ICU patient survival, data-centre failure, etc.)~\cite{fang2021pdtm, crowther2012individual}. Therefore, we choose Poisson distribution as the probability distribution to represent the neighborhood information of an event. The probability mass function of the distribution is given by the following equation:
\begin{equation}
    \small
    \mathbb{P}( k \text{ events in interval } t ) = e ^{-\lambda} \frac{\lambda^k}{k!}
\end{equation}
where $\lambda$ is the rate parameter and $t$ is normalized to 1. Usually, $\lambda$ is interpreted as expected number of events per unit time\footnote{For sufficiently large observations (>10000), $\lambda$ often becomes Gaussian mean.}. \textit{To the best of our knowledge, this is the first work that uses probability distribution to encode neighborhood information of an event node in the provenance graph.}

Let's describe the encoding process now. In ANUBIS design, we consider 4 different types of events of the OpTC dataset (\verb!process!, \verb!file!, \verb!flow!, and \verb!shell!). In other words, in the neighborhood of a newly generated event, 4 different types of events are present. Each type of event has their own Poisson distribution for a specific neighborhood\footnote{Authors of~\cite{wang2020you} claimed that given the count and variation of events in the provenance graph, it is extremely unlikely for a single multivariate distribution to capture the whole picture. We agree with that view. In our work, each event type has their own distribution.}. We calculate the difference between expected number of events and actual number of events from the distribution parameters. We store the difference in a vector, $D = [D_{process},D_{file},D_{flow}$ $,D_{shell}]$. Let us denote the expected number of events for event type $t$ as $\mathbb{E}_t$ and the actual number of events as $\mathbb{A}_t$. We formally define $D$ by the following equation:
\begin{equation}
    \small
    D_{type} = \begin{cases}
    \mathbb{E}_{process} - \mathbb{A}_{process}, & \text{when type = process} \\
    \mathbb{E}_{file} - \mathbb{A}_{file}, & \text{when type = file}\\
    \mathbb{E}_{flow} - \mathbb{A}_{flow}, & \text{when type = flow}\\
    \mathbb{E}_{shell} - \mathbb{A}_{shell}, & \text{when type = shell}
    \end{cases}
\end{equation}
\emph{The vector $D$ signifies the expected nature of the neighborhood and if there are any deviations from it so far}. Additionally, we can infer the potential events that can take place soon after the event in consideration from the distribution as well. We encode this information in a vector $P_{event}=[P_{file},P_{process},P_{flow},$ $P_{shell}]$ which is defined by Equation \ref{eq:define_p}.

\begin{equation}
\small
    P_{type} = 
    \begin{cases}
        \mathbb{P}(\text{Waiting Time}\leq \Delta t_{type}) , & \text{when type = event's type} \\
        \mathbb{P}(\text{Waiting time} > \Delta t_{type}), & \text{otherwise}
    \end{cases}
    \label{eq:define_p}
\end{equation}
The $\mathbb{P}$ in Equation \ref{eq:define_p} is defined by Equation 4 and \ref{eq:define_prob2}.
\begin{align}
    \small
    \mathbb{P}(\text{Waiting time}\leq\Delta t_{type}) & = 1 - e^{-\lambda \Delta t_{type}}\\
    \small
    \mathbb{P}(\text{Waiting time}>\Delta t_{type}) & = e^{-\lambda \Delta t_{type}}
    \label{eq:define_prob2}
\end{align}

Let's explain Equation \ref{eq:define_p}, 4, \ref{eq:define_prob2} with an example. Suppose, a parent event $e_{parent}$ is triggering a \emph{flow} type event $e_{curr}$, at time $t_{now}$ (e.g., \verb|iexplorer.exe| connecting to \verb|msftr.com|). Let's assume that the last \textit{flow} type event in the neighborhood before $e_{curr}$ was triggered at time $t_{prev}$. Therefore, waiting time for this event is $\Delta t_{flow} = t_{now}-t_{prev}$. 

We can calculate the probability of waiting \textbf{less or equal} to $\Delta t_{flow}$ for a flow event by using Equation 4. This probability is the $P_{flow}$ component of the $P_{event}$ vector described above. $P_{flow}$ signifies whether the flow event arrived at a natural interval or not. For example, $P_{flow}$ being very close to 1 or 0 means current event arrived too late or too early respectively. \emph{Essentially, probability value that we get from our Poisson encoding process is a measure of irregularity of an event.}

Note that we have to calculate the probability of other types of events too. This means calculating the $P_{process},P_{file},$ and $P_{shell}$ for the time instant of a \textit{flow} event. We assume that a parent process event can only generate one child event at a specific time instant\footnote{One can argue that in a system with multi-core processors, a process can read a file and execute a shell command at the same time by using separate processor cores. We are not considering this case in this work.}. When a specific type of event is being generated, the other type of events are \emph{still waiting} to be generated which makes their respective waiting time \textbf{greater} than $\Delta t_{type}$. This probability can be calculated by Equation \ref{eq:define_prob2}. 




\emph{The vector $D$ tells us a neighborhood's behavior so far while the vector $P_{event}$ tells us about what is going to happen in the event's neighborhood in near future. We concatenate both vectors and denote it as $E_n$. This $E_n$ gives us a near-complete picture of the neighborhood (i.e., what happened so far in the neighborhood, is it consistent with the observed Poisson distribution patterns and how the neighborhood will evolve in near future)}. Let us denote the length of $E_n$ as $d_2$. Therefore, the neighborhood encoding of an event trace of length $l$ will have the dimension $l\times d_2$. 

\subsubsection{Training Neural Network.} Consider an event trace of length $l$. The causal and contextual information encoding gives us a $l\times d_1$ matrix. The neighborhood encoding gives us a $l\times d_2$ matrix. We input these two matrices into the classifier. The classifier consists of two parallel LSTM layers and a Bayesian Neural Network (BNN). We used the ELBO loss to train the classifier which is described in the original paper proposing BNN~\cite{blundell2015weight}.

\begin{figure}[t]
    \centering
    \includegraphics[width=0.45\textwidth]{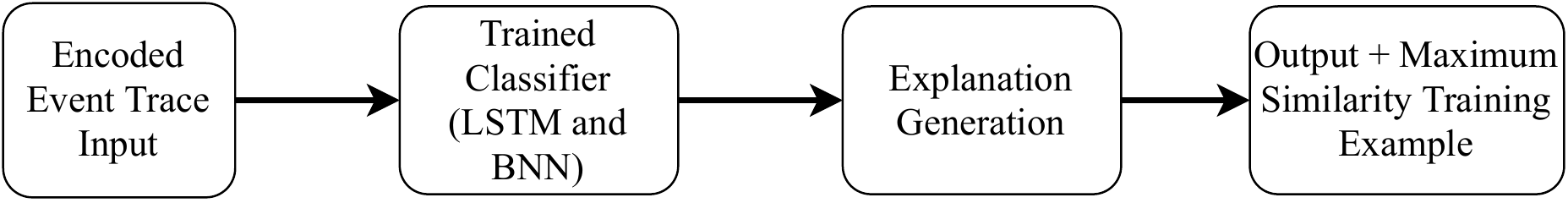}
    \caption{Block diagram showing the ANUBIS operating on encoded trace from provenance graph.}
    \label{fig:anubis_operation}
\end{figure}
\subsection{Operation Phase Design}
During the operation phase, the input of ANUBIS consists of $l$-length event traces from a streaming provenance graph. The operation phase consists of two steps. In the first step, ANUBIS makes a prediction on the input event trace (benign or APT). If the event trace is predicted as APT and the certainty level is high, ANUBIS proceeds to second step. In second step, ANUBIS creates an explanation for the prediction (similar to attack story reconstruction process). Figure~\ref{fig:anubis_operation} shows the structure of the operation phase of ANUBIS. 

In this work, we provide explanations when ANUBIS predicts an APT with high certainty. According to~\cite{threat2018fatigue}, companies get more than 17000 threat alerts per week where more than 51\% of the alerts are false positives. Due to the frequency of the alerts, only 4\% of them are properly investigated~\cite{hassan2019nodoze}. Our primary goal is to assist the cyber-threat responders by providing necessary explanation for an APT prediction to ensure trustworthy and timely investigation of incidents. Therefore, high certainty APT predictions demand higher priority in analysis. We intend to explore the explanation analysis for low certainty APT predictions in future. 

\subsubsection{ Prediction of Event Traces.} The trained classifier takes an event trace as input and makes a prediction. The prediction generates two values. The first is the prediction of the class (benign or APT). The second is the uncertainty score associated with it (i.e., standard deviation of the predictions). 

Let's illustrate the prediction process with an example. Consider a test event trace from the provenance graph. We make $k$ predictions on this event trace. Recall that our classifier (BNN) does not have specific weights and biases. Instead, it has a learned distribution of weights and biases. During each prediction, we sample weights and biases from this distribution. Therefore, in each step, we may get a different prediction result. This is illustrated in Table \ref{tab:prediction_process}. 

If the classifier is highly certain about the class of an event trace, it will predict the same class with almost similar probability in each iteration. In other words, the standard deviation of the predictions would be very low. On the contrary, if the classifier has low confidence in its prediction, it will predict different classes with varying probability in each iteration. 
This leads to a higher standard deviation of the predictions which can be seen from Table \ref{tab:prediction_process}. The low certainty prediction has 5 times more standard deviation than the high certainty prediction.

In ANUBIS design, we consider an event trace as APT if probability is greater than 0.5. If the prediction have a standard deviation less than or equal to 0.1, then it is considered as a high certainty prediction\footnote{The value 0.1 was chosen by trial and error. We initially tried with 0.05. However, the results were not great. Recall that lower standard deviation means concentrated observations. Therefore, the lower the std. dev of predictions, the better since it indicates certainty level in the prediction. In other words, no matter how many times you ask the network, it gives more or less same answer which indicates that the network has high confidence in it's prediction.}. Once ANUBIS determines the nature of the prediction, it starts creating an explanation. This is the attack story reconstruction problem which is well studied in literature~\cite{alsaheel2021atlas, pei2016hercule}.

Attack story reconstruction problem can be viewed from two perspective. The first perspective is: \textbf{What} constitutes the attack? The work in~\cite{alsaheel2021atlas} proposes a system named ATLAS that solves the problem from this perspective. ATLAS creates a sequence of events from the optimized causal graph and trains an RNN-based model to make prediction on it. If the trained model predicts an event sequence as malicious, ATLAS constructs the attack story by mapping the sequence back to their causal events. The second perspective is: \textbf{Why} this is an attack? It requires that any malicious prediction made by the system is mapped to a member of the training set that has the maximum similarity. In ANUBIS design, we adopt this perspective for explainability. \textit{To the best of our knowledge, this is the first work that leverages the second perspective to achieve attack explainability.} 
\begin{table}[t]
    \centering
    \footnotesize
    
    \caption{Illustration of predictions on test event trace . Closer to 0 means benign and closer to 1 means malicious.}
    \begin{tabular}{|c|c|c|c|c|c|c|c|}
    \hline 
    \multicolumn{4}{|c|}{High Certainty} & \multicolumn{4}{c|}{Low Certainty} \\
    \hline 
    \# & Prediction & \# & Prediction & \# & Prediction & \# & Prediction \\
    \hline 
    \hline 
    1 & 0.84 & 6 & 0.92 & 1 & 0.663 & 6 & 0.512 \\
    \hline 
    2 & 0.882 & 7 & 0.814 & 2 & 0.541 & 7 & 0.986 \\
    \hline 
    3 & 0.891 & 8 & 0.794 & 3 & 0.741 & 8 & 0.186 \\
    \hline
    4 & 0.872 & 9 & 0.914 & 4 & 0.519 & 9 & 0.616 \\
    \hline 
    5 & 0.92 & 10 & 0.886 & 5 & 0.341 & 10 & 0.781 \\
    \hline 
    \hline 
    \multicolumn{2}{|c}{mean} & \multicolumn{2}{|c|}{0.8733} & \multicolumn{2}{|c}{mean} & \multicolumn{2}{|c|}{0.5886}\\
    \hline
    \multicolumn{2}{|c}{std. dev} & \multicolumn{2}{|c|}{0.0418} & \multicolumn{2}{|c}{std. dev} & \multicolumn{2}{|c|}{0.2146}\\
    \hline
    \end{tabular}
    \label{tab:prediction_process}
    
\end{table}
\subsection{Explanation of High Certainty Malicious Prediction.} A Neural Network consists of several layers. Each layer takes a vector as input and produces a vector as output. The output is called the ``activation'' of the layer. We leverage the activation of layers to get the training event trace that is most similar with the input that generated the malicious prediction. 
\begin{table*}[t]
    \centering
    \footnotesize
    \caption{APT activities present in the OpTC dataset.}
    \begin{tabular}{|c|c|c|c|c|}
    \hline
    Vulnerability Code & Description &  Attack Vectors & Recent Attacks &Presence In Dataset\tablefootnote{Day numbering represents evaluation period. Not benign data collection period.} \\
    \hline
    \hline
    CVE-2021-30551 & Remote Code Execution and Shell Code Injection & Beacon (Cobalt Strike) & Google Chrome (2021)~\cite{chrome_apt} & Day 1 \\
    \hline
    CVE-2020-0688 & Remote Code Execution and Lateral Movement & Powershell Empire & Microsoft Exchange (2020)~\cite{cve_2020_0688} & Day 1 and 2\\
    \hline
    CVE-2019-0604 & Remote Code Execution and Credential Harvesting & Customized Mimikatz & Microsoft Sharepoint (2019)~\cite{cve_2019_0604} & Day 1 and 3 \\
    \hline
    
    \end{tabular}
    \label{tab:dataset_relevance}
\end{table*}

\begin{table}[t]
    \centering
    \footnotesize
    \caption{Characteristics of the constructed provenance graphs.}
    \scriptsize
    \begin{tabular}{|p{0.6cm}|p{0.8cm}|p{0.8cm}|p{1cm}|p{0.75cm}|p{0.75cm}|}
        \hline
        Graph & \# events\tablefootnote{We heavily pruned the graph. For example, we got rid of all the ICMP ping messages, bi-directional network flow messages and TCP 3 way connection setup messages.} & \# benign traces & \# APT traces & Avg. Benign Trace Length & Avg. APT Trace Length\\
        \hline 
        \hline
        Day 1 & 78,237,643 & 5,000,000 & 22,919 & 12.65 & 16.42 \\
        \hline
        Day 2 & 42,618,311 & 5,000,000 & 26,212 & 10.86 & 9.22\\
        \hline
        Day 3 & 59,140,814 & 5,000,000 & 8,343 & 13.41 & 7.95\\
        \hline
    \end{tabular}
    \label{tab:graph_details}
\end{table}

Let's assume that the trained BNN consists of $k$ layers $\{L^1,\cdots,L^k\}$. Consider an event trace $T_H$. Let's also assume that when we input $T_H$ in the BNN, it generates a high certainty malicious prediction. We denote the activation of the BNN layers for $T_H$ as $A_{T_H} = \{L_{out}^1(T_H),\cdots,L_{out}^k(T_H)\}$. Now, consider a malicious event trace $T_{train}$ from the ANUBIS training set . We calculate the activation of BNN layers for $T_{train}$ in the same manner and denote that as $A_{T_{train}} = \{L_{out}^1(T_{train}),\cdots, L_{out}^k(T_{train})\}$. The similarity $d_{trace}$ between the two event traces is judged by the following equation\footnote{$|a-b|_2$ is Euclidean distance between two vectors a and b.}.

\begin{equation}
    d_{trace} = A_{T_H} - A_{T_{train}} = \frac{1}{k} \sum_{i=1}^k |L_{out}^i(T_H)-L_{out}^i(T_{train})|_2
    \label{eq:similarity}
\end{equation}


\textit{Equation \ref{eq:similarity} essentially computes whether $T_H$ and $T_{train}$ is triggering similar kind of response (activation) from the BNN. The smaller value of $d_{trace}$ suggests greater similarity.} We compute $d_{trace}$ for all the malicious event traces in the ANUBIS training set. Let us denote the malicious event trace in the training set with the smallest $d_{trace}$ value as $\widehat{T_{train}}$. ANUBIS provides the cyber analyst with a prediction report of $T_H$ which contains $\widehat{T_{train}}$ as the most similar malicious event trace. This enables the cyber analyst to understand \textbf{why} $T_H$ was predicted as malicious. Moreover, this enables the analyst to understand the prediction from a deeper level (e.g., evolution of attack vector, if a previous vulnerability remained unpatched or if a previous patch failed to prevent the vulnerability and so on). We demonstrate this phenomenon in Section \ref{sec:exp_eval}.\\

\section{Experimental Evaluation}
\label{sec:exp_eval}
\subsection{Dataset}
We used the DARPA OpTC dataset to perform experimental evaluation of ANUBIS~\cite{optcdatasetieee}. This dataset contains APT activities conducted by a professional red team over the course of 7 days in a networked environment of one thousand windows endpoints. The reasons for choosing this dataset for experimental evaluation are as follows:
\begin{itemize}
    \item The APT examples that are present in this dataset have been used in recent high profile APT attacks. This is shown in Table \ref{tab:dataset_relevance}. 
    \item This is by far the largest dataset available from DARPA Transparent Computing Program~\cite{anjum2020analyzing}. Therefore, it is ideal to train and test deep learning models. 
    \item According to~\cite{fireeye2020report}, 89\% of APT malware focus specifically on windows systems. This dataset consists data collected from windows hosts only. Moreover, attackers predominantly use powershell, remote desktop protocol and windows management infrastructure to accomplish their goals. The OpTC dataset specifically contains numerous instances of all three techniques.
    \item No prior works on APT detection used the whole OpTC dataset. The only work we found which used this dataset is~\cite{cochrane2021sktree}. However, the authors used only a subset of the dataset (data from the first day of malicious activities). In this work, we used the whole dataset to construct the provenance graph. \emph{Therefore, to the best of our knowledge, this is the first attempt to benchmark an APT detection model on the OpTC dataset.}
\end{itemize}

The OpTC dataset contains multiple APT scenarios. We constructed provenance graphs for each scenario to evaluate ANUBIS. 
\subsection{Evaluation}

We evaluated ANUBIS on the DARPA OpTC dataset that contains approximately 17 billion OS-level events in total. Our evaluation is based on two criteria: prediction performance (benign/APT) and high certainty prediction explainability. Specifically, the evaluation analysis is driven by the following research questions:
\begin{enumerate}[label=Q\arabic*. , wide=0.5em,  leftmargin=*]
    \item How accurately ANUBIS can detect APT-related event traces? What is the false positive rate? (\cref{sec:detection})
    \item What are the effects of the design decisions we made for ANUBIS? (\cref{sec:anubis_design})
    \item How effective ANUBIS is compared to the other state of the art APT detection models? (\cref{sec:anubis_compare})
    \item How accurately can ANUBIS explain the predictions it makes? (\cref{sec:explanation})
    
\end{enumerate}

\subsubsection{Detection Performance}
\label{sec:detection}
We constructed three separate provenance graphs for the three APT activity days in the dataset. The details of the graphs are given in Table \ref{tab:graph_details}. It shows that there is a class imbalance in the dataset. To mitigate this, we oversample the APT traces during the ANUBIS training process. We split the dataset into training set and testing set by 80-20 ratio. The prediction result of ANUBIS is presented in Table \ref{tab:anubis_detect}. It is important to note that Table \ref{tab:anubis_detect} presents the results from the best configurations. The effect of configurations and other design decisions on prediction performance  will be discussed further in \cref{sec:anubis_design}. We can see from Table \ref{tab:anubis_detect} that ANUBIS has high prediction accuracy, precision and recall. Moreover, ANUBIS has very low false positive rate. As a result, we believe that our approach is less likely to induce ``threat fatigue'' among cyber analysts in compare to other approaches currently  in use. It is interesting to note that ANUBIS performs significantly well on day 3. Day 3 contains the scenario of ``Malicious Upgrade'' where a malicious binary is downloaded as a software update (similar to Solarwinds Sunburst APT attack). This shows that ANUBIS can detect APTs that are relevant in today's threat landscape.  
\begin{table}[t]
    \centering
    \scriptsize
    \caption{APT detection performance of ANUBIS.}
    \begin{tabular}{|c|c|c|c|c|c|c|}
    \hline
        Graph & Accuracy & Precision & Recall & F-score & FPR \tablefootnote{False Positive Rate} & \# False Positive \\
        \hline
        \hline
        Day 1 & 0.99 & 0.99 & 1.00 & 0.998 & 0.001 & 147\\
        \hline
        Day 2 & 0.99 & 0.98 & 1.00 & 0.989 & 0.007 & 235\\
        \hline
        Day 3 & 1.00 & 1.00 & 0.99 & 1.00 & 0.000 & 12\\
        \hline
        \hline
        Avg. & 0.993 & 0.99 & 1.00 & 0.996 & 0.003 & 131.33\\
        \hline
    \end{tabular}
    
    \label{tab:anubis_detect}
\end{table}
\subsubsection{Effect of Design Decisions}
\label{sec:anubis_design}
We identify two aspects of our design that can affect the detection performance of ANUBIS. The first one is the length of the event trace and the second one is the dimension of the neighborhood encoding. We present our comparative analysis for both of them.
\paragraph{Length of Event Trace.}
The length of the event trace essentially determines the amount of information we are considering for classification. We experimented with trace length $l=\{2,4,6\}$. The result is presented in Table \ref{tab:anubis_trace_length}. Note that the best results are not concentrated for a specific value of $l$. For $l=2$, the results are less attractive than $l=\{4,6\}$. An event trace of length 2 has the contextual and neighborhood information from the parent and the event itself. On the other hand an event trace of length 4 or 6 has additional information regarding parents of it's parent event. This richness of information allows ANUBIS to perform better classification on event traces of length 4 and 6. We did not experiment on trace length more than 6 as improvement of metrics between trace length 4 and 6 are marginal. However, this can be easily done if required in the operation phase. Also, in the next subsection, we will show how our novel neighborhood encoding increases the detection performance greatly even when $l$ is small. 

\begin{table}[t]
    
    \centering
    \scriptsize
    \caption{Event trace length effect in operation phase.}
    \begin{tabular}{|c|c|c|c|c|c|c|c|c|}
    \hline
    \multirow{2}{*}{Graph} &  \multirow{2}{*}{$l$} & \multicolumn{5}{c|}{Metrics} & \multicolumn{2}{c|}{System Memory (MB)} \\
    \cline{3-9}
         &  & Accuracy & Precision & Recall & F-score & FPR & Graph & Model\\
    \hline
    \hline

    \multirow{3}{*}{Day 1} & 2 & 0.979 & 0.932 & 0.961 & 0.965 & 0.021 & 98.6 & 5.8\\  
    \cline{3-9}
    & 4 & \textbf{0.993} & \textbf{0.997} & \textbf{1.00} & \textbf{0.998} & \textbf{0.001} & 351.4 & 6.1 \\  
    \cline{3-9}
    & 6 & 0.984 & 0.991 & \textbf{1.00} & 0.985 & 0.002 & 572.9 & 6.3\\  
    \hline
    \hline
    
    \multirow{3}{*}{Day 2} & 2 & 0.964 & 0.902 & 0.944 & 0.922 & 0.034 & 72.1 & 5.8\\  
    \cline{3-9}
    & 4 & \textbf{0.990} & 0.977 & \textbf{1.00} & 0.988 & 0.011 & 297.1 & 6.1\\  
    \cline{3-9}
    & 6 & 0.989 & \textbf{0.983} & \textbf{1.00} & \textbf{0.989} & \textbf{0.007} & 472.4 & 6.3 \\
    \hline
    \hline
    
    \multirow{3}{*}{Day 3} & 2 & 0.939 & 0.915 & 0.927 & 0.921 & 0.029 & 77.4 & 5.8\\  
    \cline{3-9}
    & 4 & 0.997 & 0.977 & \textbf{1.00} & 0.988 & 0.004 & 313.8 & 6.1\\  
    \cline{3-9}
    & 6 & \textbf{1.00} & \textbf{1.00} & 0.996 & \textbf{1.00} & \textbf{0.00} & 533.6 & 6.3\\
    \hline
    
    \end{tabular}
    
    \label{tab:anubis_trace_length}
\end{table}

The length of the event trace determines the size of the provenance graph required to be stored during the operation phase. For instance, if we consider event trace length 4, at least 3 previous parent events are required to be present in the graph for any event trace. In the system memory column of Table \ref{tab:anubis_trace_length}, we present the size (in Megabytes) of the in-memory provenance graph that is required for the operation phase. As $l$ increases, the provenance graph size monotonically increases. Same goes for the model size (BNN and RNN layers) during runtime. However, it is important to note that model size increase is very small. This is because, when the trace event length increases, only the input layer of the model architecture changes. In our analysis, we are not considering the overhead of system level provenance event generation. Practical whole-system provenance mechanisms are present in the literature ~\cite{Hi2012fi, pasquier2017practical} and ANUBIS will leverage them to construct the graph in the operation phase. 

\paragraph{The Dimension of Neighborhood Encoding.} The dimension $d_2$ of the neighborhood encoding also plays a role in the operation phase performance. In the neighborhood encoding design, we concatenated $D$ and $P_{event}$ vectors. Both of them have a component for a specific event type. We experimented with different number of event types to get different dimensions of the neighborhood encoding. For example, if we only consider \verb|process| events, then $d_2 = 2$. If both \verb|process| and \verb|file| are considered, then $d_2=4$ and so on. In our design, we used 4 types of events which resulted in $d_2= 8$. The prediction results for neighborhood encoding with different $d_2$ values for Day 1 are presented in Table ~\ref{tab:anubis_n_encoding}. The results for Day 2 and 3 graph are omitted for space issues. 

\begin{table}[t]
    \centering
    \scriptsize
    \caption{Neighborhood encoding length effect in operation phase}
    \begin{tabular}{|c|c|c|c|c|c|c|}
    \hline
    Graph & $d_2$ & Precision & Recall & F-score & Time (s) & Memory (MB)\\
    \hline
    \hline
    \multirow{4}{*}{Day 1 ($l=4$)} & 2 & 0.754 & 0.782 & 0.768 & 0.39 & 56.2\\
    \cline{2-7} 
    & 4 & 0.882 & 0.794 & 0.836 & 0.41 & 144.4\\
    \cline{2-7}
    & 6 & 0.914 & 0.929 & 0.921 & 0.83 & 346.1\\
    \cline{2-7}
    & 8 & 0.997 & 1.00 & 0.998 & 0.97 & 572.9 \\
    \hline
    
    \end{tabular}
    \label{tab:anubis_n_encoding}
    
\end{table}

Table \ref{tab:anubis_n_encoding} gives us some interesting insights into the design. As the length of neighborhood encoding increases, classification performance increases greatly. This can be attributed to the fact that granular neighborhood encoding allows ANUBIS to learn about distributions more accurately. In other words, the less we limit information regarding neighborhood, the better ANUBIS learns about potential discrepancies in an event's neighborhood. Based on this observation, it can be argued that the encoding process will be better if we consider more granularity. For example, instead of a single distribution for \verb|process|, we should consider different distributions for \verb|process-create|, \verb|process-open| and \verb|process-terminate|. For \verb|file|, we should consider \verb|file-create|,\verb|file-open|,\verb|file-modify| and so on. This is accurate in theory. However, the practicality of this approach is invalidated by observing the time and memory column of Table \ref{tab:anubis_n_encoding}. The time column represents the average time required to calculate neighborhood of 100 event traces. It increases monotonically with $d_2$. Therefore, if we break up the current event types into more granular event types based on their actions, it will further increase the time needed for neighborhood encoding. This can potentially be a performance issue in real world systems. For example, consider ANUBIS operating on enterprise hosts that generate a lot of OS-level events (e.g., DNS servers). For a granular neighborhood encoding, the required time for encoding will be higher which can lead to a backlog of events waiting to be encoded and sent to classifier for prediction. Additionally, it will also result in a bigger provenance graph during runtime. This is demonstrated in the memory column of Table \ref{tab:anubis_n_encoding}. The memory footprint of the graph increases monotonically with $d_2$ since it is storing lengthier floating point vectors per node. We carefully designed ANUBIS to be not very resource hungry and our choice of granularity for neighborhood encoding is optimal for our experiments.   

\begin{table}[t]
    \centering
    \scriptsize
    \caption{Summary of APT detection models in literature}
    \begin{tabular}{|p{1.35cm}|p{1.4cm}|p{1.35cm}|p{0.35cm}|p{0.35cm}|p{0.35cm}|p{0.75cm}|}
    \hline
    Model & Method & Dataset & Acc. & Prec. & Rec. & F-score \\
    \hline 
    \hline
    Unicorn~\cite{han2020unicorn} & Unsupervised & DARPA TC3 & 0.99 & 0.98 & 1 & 0.99 \\
    \hline
    StreamSpot~\cite{manzoor2016fast} & Unsupervised & Own Dataset & 0.66 & 0.74 & N/A & N/A \\ 
    \hline 
    Provdetector~\cite{wang2020you} & Unsupervised & Own Dataset & N/A & 0.959 & 1 & 0.978\\
    \hline
    Holmes~\cite{Milajerdi_2019} & Edge Matching & DARPA TC3 & N/A & 0.99 & 0.99 & 0.99 \\
    \hline 
    Poirot~\cite{milajerdi2019poirot} & Graph Matching & DARPA TC3 & N/A & 0.99 & 0.99 & 0.99 \\
    \hline
    Atlas~\cite{alsaheel2021atlas} & Supervised & Own Dataset & N/A & 0.998 & 0.998 & 0.998\\
    \hline
    \hline
    \textbf{Anubis} & \textbf{Supervised} & \textbf{DARPA OpTC} & \textbf{0.993} & \textbf{0.99} & \textbf{1} & \textbf{0.996}\\
    \hline
    \end{tabular}
    
    \label{tab:other_works}
\end{table}

\subsubsection{Evaluation with respect to State-of-the-art Models}
\label{sec:anubis_compare}
Table \ref{tab:other_works} gives a non-exhaustive summary of the APT detection models available in the literature. \emph{It is important to understand that direct comparison with these models is not possible due to a lot of factors}. Some of the prior models used older or custom datasets. Some of them used unsupervised models for detecting APTs. We report the best results from these models. We did not include results from models~\cite{hassan2019nodoze, milajerdi2018propatrol, sun2018zepro} that used different metrics than the ones we reported.  
\begin{table}[t]
    \centering
    \scriptsize
    \caption{APT Prediction By Explainability}
    \begin{tabular}[width = 0.45\textwidth]{|p{0.5cm}|p{1.15cm}|p{1.45cm}|p{1.5cm}|p{0.8cm}|}
    \hline
    Graph  & \# Predictions & Low uncertainty & High uncertainty& Incorrect \\
    \hline
    \hline
    Day 1 & 6876 & 6523 & 353 & 31 \\
    \hline
    Day 2 & 7863 & 7805 & 58 & 94\\
    \hline 
    Day 3 & 2503 & 2466 & 37 & 12 \\
    \hline
    \end{tabular}
    
    \label{tab:prediction_confidence_report}
\end{table}

From Table \ref{tab:other_works}, it is evident that ANUBIS's performance is comparable to the state of the art APT detection models. ANUBIS achieves 99.3\% accuracy with F-score of 0.996. UNICORN, Holmes and Poirot achieve similar accuracy and F-score on DARPA TC3 dataset. Note that DARPA TC3 is a smaller dataset (approximately 400 million events) with attacks on UNIX, BSD and Android ecosystems~\cite{han2020unicorn} while DARPA OpTC is newer and orders of magnitude bigger in terms of events (approximately 17 billion events). Therefore, the OpTC dataset has much lower signal to noise ratio compared to the TC3 dataset. In other words, the OpTC dataset is an extreme example of ``needle in a haystack" scenario when compared with the TC3 dataset. 


ANUBIS achieves 2.28\% false positive rate which is very low (Table~\ref{tab:anubis_detect}) when loosely compared to similar models. For example, UNICORN reported 8\% false positive rate in their experiment on the Streamspot dataset~\cite{manzoor2016fast}. UNICORN did not report their false positive rate on the TC3 dataset. Moreover, Holmes and Poirot both reported ``very low" false positive rate on TC3 without quantifying any value. Streamspot did not report any false positive rate. However, given that Streamspot has accuracy and precision below 80\%, it is reasonable to assume that they have higher false positive rate than all other models in Table~\ref{tab:other_works}.

\subsubsection{Prediction Explainability}
\label{sec:explanation}
\begin{figure}[t]
    \centering
    \includegraphics[width=0.45\textwidth]{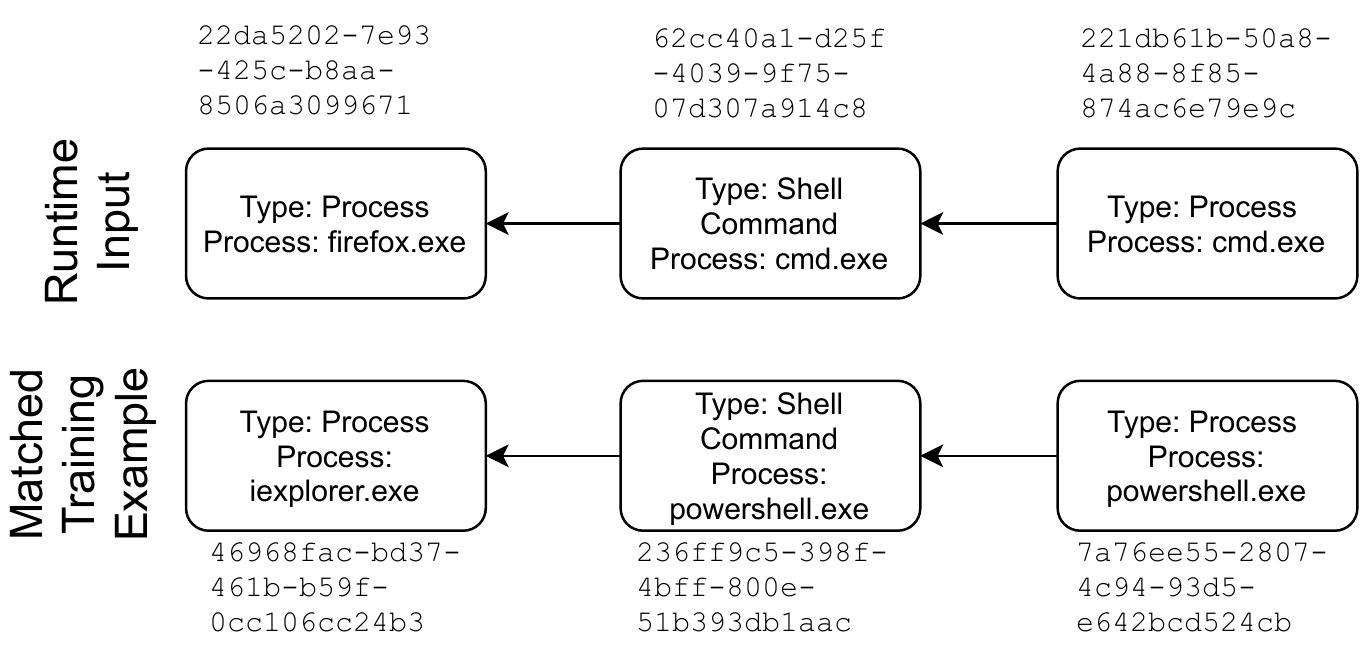}
    \caption{Explaining prediction with low Uncertainty}
    \label{fig:low_certainty_prediction}
\end{figure}

In the design section, we described how ANUBIS handles the explanation of low uncertainty malicious predictions. Table \ref{tab:prediction_confidence_report} presents the malicious test set classifications by their uncertainty level. Day 1 has the maximum number of high uncertainty prediction whereas day 3 had the least amount of high uncertainty predictions. In Figure \ref{fig:low_certainty_prediction}, we show a low uncertainty prediction. ANUBIS predicted the event trace as malicious with probability 0.91 and standard deviation 0.024. This input event trace was matched with the training example in Figure \ref{fig:low_certainty_prediction}. At a first glance, two event traces do not offer any resemblance. However, detailed analysis reveals that ANUBIS learned about higher level patterns. 


In the training example, the first powershell process launches a second powershell process. The first one injects the second powershell process with a shellcode. The shellcode contains this statement: \begin{Verbatim}[breaklines=true]
    SySteM.NEt.SeRviCePoInTMaNaGeR]:: EXPEct100COnTiNUe=0;$WC=NEW-ObJecT  SyStEm.NEt.WebClIEnT;$u='Mozilla/5.0 (Windows NT 6.1; WOW64; Trident/7.0; rv:11.0)
\end{Verbatim}
This code is using the Windows web client functionality to reach a web domain for malicious package download. However, the system does not have Mozilla Firefox installed. That is why \verb|iexplorer.exe| is launched. The pattern we see here is multiple shell command process are injected with shellcode which launches the network process for malicious domain visit. The input event trace follows the same pattern but with different shell command clients and internet browser. The input event trace injects shellcode in \verb|cmd.exe| and visits the malicious domain with \verb|firefox.exe|. ANUBIS understands that this input event trace has the same high level structure as the training example in Figure \ref{fig:low_certainty_prediction}. The cyber analyst gets a comprehensive report from ANUBIS which includes the prediction and attack story based on the matched training example.

\section{Discussion and Limitations}
ANUBIS shares a few limitations with other APT detection models in the literature. Firstly, ANUBIS is trained on a dataset that contains data from windows hosts only. As a result,  ANUBIS does not cover all kinds of hosts in an enterprise scenario. We intend to test ANUBIS on different os-level provenance graphs in future. 

Another issue is that ANUBIS requires high volume of data for training since it depends  on deep learning layers. This might prove challenging for organizations with less computing power. Therefore, ANUBIS is more suited for use in an enterprise scenario. However, we also notice that ANUBIS performed extremely well in a homogeneous enterprise environment. The DARPA OpTC dataset contained data from 1000 windows hosts with similar workload. If the workload varies in hosts (e.g., workstations with different users), ANUBIS may require more training time and data to capture that pattern. In general, APT detection in heterogeneous hosts is a difficult problem~\cite{han2020unicorn}. We intend to explore the applicability of ANUBIS in hosts with diverse workloads in future. 

ANUBIS requires periodic re-training to function effectively against new APT attacks. For example, an analyst can judge the quality of high uncertainty event traces and can decide whether to use them for re-training. This periodic re-training ensures that the next time ANUBIS sees such an event trace it will be highly confident about the classification. However, this also opens up an attack vector for the adversary. An adversary with knowledge of the system may try to get malicious event traces into benign re-training dataset. This will cause ANUBIS to classify malicious traces as benign with high confidence and eventually undermining the security of the enterprise network. Therefore, the re-training data need to be selected and sanitized carefully before training ANUBIS. Moreover, the neural network parameters of ANUBIS also require to be tuned in order to accommodate new training data (e.g., more neurons per layer, more layers). In our future work, we intend to address the difficulties in re-training ANUBIS with new data.

\section{Related Works}
\label{sec:lit_review}
This work lies in the intersection of graph-based anomaly detection and attack story re-construction.

\noindent \textbf{Graph-Based Anomaly Detection.} There is a large body of work in the context of threat and anomaly detection based on graph~\cite{akoglu2015graph}. UNICORN~\cite{han2020unicorn} follows an unsupervised machine learning approach to analyze a streaming provenance graph for detecting APTs. It periodically computes a histogram sketch of the provenance graph which allows it to model benign system behavior. The approach is sensitive to sharp change in system behavior and can result in increase of false positive predictions. PROVDETECTOR~\cite{wang2020you} extracts rare execution paths from the provenance graph. These rare paths are vectorized using \verb|doc2vec| embedding model and then local outlier factor algorithm is used to classify the benign and malicious events. Both these approaches are unsupervised while ANUBIS is supervised. Moreover, none of these approaches offer causality analysis on the predictions. ANUBIS performs extensive causality analysis based on the confidence level of the prediction. 

StreamSpot~\cite{manzoor2016fast} uses feature engineering from streaming information flow graphs to detect malicious activities. The feature engineering process extracts local graph structures from a single snapshot of the provenance graph. However, this approach is ineffective against APT which sometimes have campaign length of multiple months or years~\cite{fireeye2020report}. Moreover, StreamSpot triggers a large number of false alarms which can cause threat-fatigue and result in preemptive dismissal of alerts~\cite{hassan2019nodoze}. FRAPpuccino~\cite{han2017frappuccino} is another approach for provenance graph-based APT detection. It performs a sliding window analysis which results in a feature vector based on counts. A model is trained on these feature vectors. The model represents normal execution state of the system. Any deviation from the model is considered as anomaly. Using count for feature engineering is a major weakness of FRAPpuccino. We have shown in Section \ref{subsec:training} that count is a weaker density estimate to model the provenance graph neighborhood. 

ZePro~\cite{sun2018zepro} is the first work that uses probabilistic approach for anomaly detection. ZePro differs from ANUBIS in many ways. Firstly, ZePro uses Probability Network (shallow learning model) while ANUBIS uses BNN (deep learning model). Secondly, ZePro works on System Object Dependency Graph (graph has cycles) where ANUBIS works on provenance graph (directed acyclic graph). Finally, ZePro solves zero-day attack path identification problem. On the other hand, ANUBIS solves the APT detection and explainability problem. APTs may or may not have zero-day exploits in them (although having zero-day is most common for APT). Therefore, ZePro solves a subset of the problem which ANUBIS solves. 

Gao \textit{et al.} designed a complex query language SaQL~\cite{gao2018saql} that can query on a streaming provenance data. Based on the queries, the system leverages multiple anomaly detection model (rule-based, time-series-based, invariant-based, outlier-based anomaly detection models). The detection system aggregates data from multiple hosts and has showed promising performance in enterprise network settings. However, the rule-based matching still requires expert domain knowledge. ANUBIS does not require any expert domain knowledge during training or operation phase. Barre \textit{et al.} used random forest classifier to detect APT by feature engineering from provenance graph~\cite{barre2019mining}. They used features that are likely to be important in APT detection (e.g., timespan, edge count, and type of event count). However, the system only had $\approx$50\% detection rate. This exemplifies that feature engineering has limited value in APT detection unless execution context is considered. ANUBIS takes context into account by encoding causal and neighborhood information in the training inputs. \\

\noindent \textbf{Attack Story Reconstruction and Explainability.} One of the earliest works to leverage provenance graphs for attack story reconstruction is BackTracker~\cite{king2003backtracking}. This work leverages the provenance graph to detect the entry point of an intrusion. BackTracker is farther improved in PriorTracker~\cite{liu2018towards} which includes both forward and backward causality analysis. Both of these works perform causality analysis by graph traversal. On the contrary, ANUBIS takes a fundamentally different approach as it performs causality analysis by matching the event trace with the most similar event trace in the training set. 

ProPatrol~\cite{milajerdi2018propatrol} approaches attack story reconstruction problem by compartmentalizing high level activity. This mitigates the dependence explosion problem which hinders forensic analysis of APT attacks. HOLMES~\cite{Milajerdi_2019} is another work that focuses on attack story reconstruction by high-level graph abstraction. HOLMES uses MITRE TTPs~\cite{alexander2020mitre} to correlate provenance graph events and generate a high level overview of the attack story. POIROT~\cite{milajerdi2019poirot} shares a similar strategy for attack story reconstruction. POIROT constructs a query graph from available cyber threat intelligence and tries to align the query graph with the provenance graph in order to detect APT attacks. POIROT demonstrated strong performance in offline pattern matching, however, constructing a query graph from attack description is difficult and requires significant domain expertise. ANUBIS design does not require any domain knowledge. The causality and explainability provided by ANUBIS comes from the training data.

ATLAS~\cite{alsaheel2021atlas} uses a sequence-based deep learning model for attack story reconstruction. Once ATLAS predicts an input sequence as malicious, it maps the sequence back to it's original events. The causality between the events are connected and the analyst can see the attack story. While this describes ``This is \textbf{how} APT is attacking'', it does not tell \textbf{why} the model thinks this sequence of events represent an APT attack. ANUBIS takes an event trace as input which is similar to the input sequence of ATLAS. Therefore, ANUBIS can perform the same attack reconstruction that ATLAS does. Additionally, ANUBIS also explains \textbf{why} the model thinks this as a malicious event trace. This makes ANUBIS a superior information source for attack explanation and reconstruction. 


\section{Conclusion}
In this work, we developed ANUBIS, a supervised deep learning framework for detecting advanced persistent threats from system provenance graphs. ANUBIS also explains the reasoning behind its prediction by matching a detected APT trace with similar traces in the training set. We prioritized explainability over storytelling to solve the attack story reconstruction problem since we believe this will help an analyst immensely. We also proposed a novel graph neighborhood encoding method using Poisson distribution that we showed is highly effective. We benchmarked our framework on the recently published DARPA OpTC APT dataset. ANUBIS achieved accuracy over 99\% and a low false positive rate of 2.8\%. ANUBIS also learned about high-level APT patterns to accurately explain the attack story to the cyber analyst. In the future, we will test ANUBIS on different datasets and develop new methods for attack story explanation. 

\bibliographystyle{ACM-Reference-Format}
\bibliography{acmart}
\end{document}